\documentclass[sigconf]{acmart}

\usepackage{booktabs} 
\usepackage{listings}
\usepackage{multirow}
\usepackage{balance}
\PassOptionsToPackage{hypen}{url}
\usepackage{hyperref}
\PassOptionsToPackage{hidelinks}{hyperref}
\hypersetup{breaklinks=true}
\newcommand{\squishlist}{
   \begin{list}{$\bullet$}
    { \setlength{\itemsep}{0pt}      \setlength{\parsep}{3pt}
      \setlength{\topsep}{3pt}       \setlength{\partopsep}{0pt}
      \setlength{\leftmargin}{3.5mm} \setlength{\labelwidth}{1em}
      \setlength{\labelsep}{0.5em} } }

\newcommand{\squishlisttwo}{
   \begin{list}{$\bullet$}
    { \setlength{\itemsep}{0pt}    \setlength{\parsep}{0pt}
      \setlength{\topsep}{0pt}     \setlength{\partopsep}{0pt}
      \setlength{\leftmargin}{2em} \setlength{\labelwidth}{1.5em}
      \setlength{\labelsep}{0.5em} } }

\newcommand{\squishend}{
    \end{list}  }

\newcommand{\code}[1]{{\tt #1}}

\newcommand{\refsec}[1]{Sec.~\ref{#1}}

\definecolor{darkgray}{rgb}{.4,.4,.4}
\newcommand\pythonstyle{\lstset{
    language=Python,
    aboveskip={1.3\baselineskip},
    basicstyle=\scriptsize\ttfamily\linespread{4},
    keywordstyle=\color[HTML]{228B22}\bfseries,
    commentstyle=\color[rgb]{0.127,0.427,0.514}\ttfamily\itshape,
    frame=single,
    identifierstyle=\color{black},
    stringstyle=\color[rgb]{0.639,0.082,0.082}\ttfamily,
    numbers=left,
    numberstyle=\tiny,
    upquote=true,
    literate=%
       *{0}{{{\color{darkgray}0}}}1
        {1}{{{\color{darkgray}1}}}1
        {2}{{{\color{darkgray}2}}}1
        {3}{{{\color{darkgray}3}}}1
        {4}{{{\color{darkgray}4}}}1
        {5}{{{\color{darkgray}5}}}1
        {6}{{{\color{darkgray}6}}}1
        {7}{{{\color{darkgray}7}}}1
        {8}{{{\color{darkgray}8}}}1
        {9}{{{\color{darkgray}9}}}1
}}
\lstnewenvironment{python}[1][]{\pythonstyle\lstset{#1}}{}

\copyrightyear{2017}
\acmYear{2017}
\setcopyright{rightsretained}
\acmConference{KDD'17}{}{August 13--17, 2017, Halifax, NS, Canada.} \acmPrice{}
\acmDOI{http://dx.doi.org/10.1145/3097983.3098171}
\acmISBN{978-1-4503-4887-4/17/08}

\balance
\begin{document}
\title{TensorFlow Estimators: Managing Simplicity vs. Flexibility in High-Level Machine Learning Frameworks}

\author[Cheng et al.]{Heng-Tze Cheng$^\dagger$ \and Zakaria Haque$^\dagger$ \and Lichan Hong$^\dagger$ \and Mustafa Ispir$^\dagger$ \and Clemens Mewald$^\dagger$\footnotemark[1] \and Illia Polosukhin$^\dagger$ \and Georgios Roumpos$^\dagger$ \and D Sculley$^\dagger$ \and Jamie Smith$^\dagger$ \and David Soergel$^\dagger$ \and Yuan Tang$^\ddagger$ \and Philipp Tucker$^\dagger$ \and Martin Wicke$^\dagger$\footnotemark[1] \and Cassandra Xia$^\dagger$ \and Jianwei Xie$^\dagger$}
\affiliation{
  \institution{$^\dagger$Google, Inc.\qquad$^\ddagger$Uptake Technologies, Inc.}\vspace*{3mm}
  \authornote{Corresponding authors: \{clemensm,wicke\}@google.com}
}

\renewcommand{\shorttitle}{TensorFlow Estimators}
\renewcommand{\shortauthors}{Cheng et al.}

\begin{abstract}
We present a framework for specifying, training, evaluating, and deploying
machine learning models. Our focus is on simplifying cutting edge machine
learning for practitioners in order to bring such technologies into production.
Recognizing the fast evolution of the field of deep learning, we make no attempt
to capture the design space of all possible model architectures in a domain-
specific language (DSL) or similar configuration language. We allow users to
write code to define their models, but provide abstractions that guide
developers to write models in ways conducive to productionization. We also
provide a unifying \code{Estimator} interface, making it possible to write
downstream infrastructure (e.g. distributed training, hyperparameter tuning)
independent of the model implementation.

We balance the competing demands for flexibility and simplicity by offering APIs
at different levels of abstraction, making common model architectures available
out of the box, while providing a library of utilities designed to speed up
experimentation with model architectures. To make out of the box models flexible
and usable across a wide range of problems, these canned \code{Estimator}s are
parameterized not only over traditional hyperparameters, but also using
\textit{feature columns}, a declarative specification describing how to
interpret input data.

We discuss our experience in using this framework in research and production
environments, and show the impact on code health, maintainability, and
development speed.

\end{abstract}

%
%



\maketitle

\section{Introduction}
Machine learning, and in particular, deep learning, is a field of growing importance. With the deployment of large GPU clusters in datacenters and cloud computing services, it is now possible to apply these methods not only in theory, but integrate them successfully into production systems.

Engineers working on production systems have only recently gained the ability to apply advanced machine learning, driven in large part by the availability of machine learning frameworks that implement the lower level numerical computations in efficient ways and allow engineers to focus on application-specific logic (see e.g., \cite{chainer, tensorflow, torch, dl4j, caffe, theano, mxnet, sklearn, mllib, xgboost, dsstne, cntk}). However, the huge amounts of data involved in training, especially for deep learning models, as well as the complications of running high intensity computations efficiently on heterogeneous and distributed systems, has prevented the most advanced methods from being widely adopted in production.

As the field of deep learning is still young and developing fast, any framework hoping to remain relevant must be expressive enough to not only represent today's model architectures, but also next year's. If the framework is to be used for experimentation with model architectures (most serious product work requires at least some experimentation), it is also crucial to offer the flexibility to change details of models without having to change components that are deeply embedded, and which have a highly optimized, low level implementation.

There is a natural tension between such flexibility on the one hand, and simplicity and robustness on the other hand. We use simplicity in a broad sense: From a practitioner's point of view, implementing models should not require fundamentally new skills, assuming that the model architecture is known. Experimenting with model features should be transparent, and should not require deep insights into the inner workings of the framework used to implement the model. We talk of robustness both as a quality of the software development process, as well as a quality of the resulting software. We call a framework robust if it is easy to write correct and high-quality software using it, but hard to write broken or poorly performing software. A framework which nudges the developer to use best practices, and which makes it hard to ``shoot yourself in the foot'' is robust.

Because of the need to keep up with and enable research, many deep learning frameworks value flexibility above all else (e.g., \cite{chainer, tensorflow, torch}). They achieve this flexibility by providing relatively low-level primitive operations (e.g., \code{matmul}, \code{add}, \code{tanh}), and require the user to write code in a regular programming language in order to specify their model. To simplify life for their users and speed up development, these frameworks often provide some higher level components, such as layers (e.g., a fully connected neural network layer with an optional activation function). Development in a fully-fledged programming language is inherently dangerous. Working at a low level can also lead to a lot of code duplication, with the software maintenance headaches that come with that.

On the other end of the spectrum are systems which use a DSL to describe the model architecture (e.g., \cite{cntk, caffe, dsstne, distbelief}). Such systems are more likely to be geared for specific production use cases. They can make common cases very simple to implement (the most common models may even be built-in primitives). Their higher level of abstraction allows these frameworks to make optimizations that are inaccessible to their more flexible peers. They are also robust: users are strongly guided towards model architectures that work, and it is hard to write down models that are fundamentally broken. Apart from the lack of flexibility when it comes to new model types and architectures, these DSL based systems can be hard to maintain in the face of an inexorably advancing body of new research. Adding more and more primitives to a DSL, or adding more and more options to existing primitives can be fatal. Google's own experience with such a system \cite{distbelief} prompted the development of TensorFlow \cite{tensorflow}.

TensorFlow is an open source software library for machine learning, and especially deep learning. It represents computation as a generalized data flow graph. The graph is first built, and then executed separately from graph construction. Operations such as \code{mul}, \code{add}, etc., are represented as nodes in the graph. Edges represent the data flowing between nodes as a \code{Tensor} containing a multi-dimensional array. In the following, we use \textit{op} and \code{Tensor} interchangeably to denote a node in the graph (op) and the output that is created when the node is executed. Most ops are stateless tensor-in-tensor-out functions. State is represented in the graph as \code{Variable}s, special stateful ops. Users can assign ops and variables to any device. A device can be a CPU, GPU, TPU, and can live on the local machine or a remote TensorFlow server. TensorFlow then seamlessly handles communication between these devices. This is one of the most powerful aspects of TensorFlow, and we rely on it heavily to enable scaling models from a single machine to datacenter-scale.

The framework described in this paper is implemented on top of TensorFlow\footnote{While we hope that our description of the features in this paper is largely self-contained, basic familiarity with TensorFlow will give valuable context to the reader.}, and has been made available as part of the TensorFlow open-source project. Faced with competing demands, our goal is to provide users with utilities that simplify common use cases while still allowing access to the full generality of TensorFlow. Consequently, we do not attempt to capture the design space of machine learning algorithms in a DSL. Instead, we offer a harness which removes boilerplate by providing best practice implementations of common code patterns. The components we provide are reusable, and integration points for users are strategically placed to encourage reusable user code. The user configuration is performed by writing regular TensorFlow code, but a number of lower level TensorFlow concepts are safely encapsulated and users do not have to reason about them, eliminating a source of common problems. 

Some of the lower level components such as layers are closely related in similar frameworks aimed at simplifying model construction \cite{keras, blocksandfuel, lasagne, slim}.

The highest level object in our framework is an \code{Estimator}, which provides an interface similar to that of Scikit-learn ~\cite{sklearn}, with some adaptations to simplify productionization. Scikit-learn has been used in a large number of small to medium scale machine learning tasks. Using a widely known interface allows practitioners who are not specialists in TensorFlow to start working productively immediately.

In the remainder of the paper, we will first discuss the overall design of our framework (\refsec{overview}), before describing in detail all major components (\refsec{components}) and our mechanisms for distributed computations (\refsec{distributed}). We then discuss case studies and show experimental results (\refsec{case}).
\section{Design Overview}\label{overview}
The design of our framework is guided by the overarching principle that users should be led to best practices, without having to abandon established idioms wherever this is possible. Because our framework is built on TensorFlow, we inherit a number of common design patterns: there is a preference for functions and closures over objects, wherever such closures are sufficient; callbacks are common. Our layer design is informed by the underlying TensorFlow style: our layer functions are also tensor-in-tensor-out operations. These preferences are stylistic in nature and have no impact on the performance or expressivity of the framework, but they allow users to easily transition if they are used to working with TensorFlow.

Because one of the greatest strengths of TensorFlow is its flexibility, it is crucial for us to not restrict what users can accomplish. While we provide guides that nudge people to best practices, we provide escape hatches and extension points that allow users to use the full power of TensorFlow whenever they need to.

Our requirements include simplifying model building in general, offering a harness that encourages best practices and guides users to a production-ready implementation, as well as implementing the most common types of machine learning model architectures, and providing an interface for developers of downstream frameworks and infrastructure. We are therefore dealing with three distinct (but not necessarily disjoint) classes of users: users who want to build custom machine learning models, users who want to use common models, and users who want to build infrastructure using the concept of a model, but without knowledge of the specifics.

These user classes inform the high level structure of our framework. At the heart is the \code{Estimator} class (see Section \ref{estimator}). Its interface (modeled after the eponymous concept in Scikit-learn \cite{sklearn}) provides an abstraction for a machine learning model, detailed enough to allow for downstream infrastructure to be written, but general enough to not constrain the type of model represented by an \code{Estimator}. \code{Estimator}s are given input by a user-defined input function. We provide implementations for common types of inputs (e.g., input from numpy \cite{numpy}).

The \code{Estimator} itself is configured using the \code{model\_fn}, a function which builds a TensorFlow graph and returns the information necessary to train a model, evaluate it, and predict with it. Users writing custom \code{Estimator}s only have to implement this function. It is possible, and in fact, common, that \code{model\_fn} contains regular TensorFlow code that does not use any other component of our framework. This is often the case because existing models are being adapted or converted to be implemented in terms of an \code{Estimator}. We do provide a number of utilities to simplify building models, which can be used independently of \code{Estimator} (see \refsec{layers}). This mutual independence of the abstraction layers is an important feature of our design, as it enables users to choose freely the level of abstraction best suited for the problem at hand.

It is worth noting that an \code{Estimator} can be constructed from a Keras \code{Model}. Users of this compatibility feature cannot use all features of Estimator (in particular, one cannot specify a separate inference graph with this method), but it is nevertheless useful for comparisons, and to use existing models inside downstream infrastructure (such as \cite{tfx}).

We also provide a number of \code{Estimator} implementations for common machine learning algorithms, which we called Canned \code{Estimator}s (these are subclasses of \code{Estimator}, see Section \ref{cannedestimators}). In our implementations, we use the same mechanisms that a user who writes a custom model would use. This ensures that we are users of our own framework. To make them useful for a wide variety of problems, canned \code{Estimator}s expose a number of configuration options, the most important of which is the ability to specify input structure using feature columns.
\section{Components}\label{components}
In this section we will describe in detail the various components that make up our framework and their relationships. We start with layers, lower-level utilities that can be used independently of \code{Estimator}, before discussing various aspects of \code{Estimator} itself.

\subsection{Layers}\label{layers}
One of the advantages of Deep Learning is that common model architectures are built up from composable parts. For deep neural networks, the smallest of these components are called network \textit{layers}, and we have adopted this name even though the concept is more widely applicable. A layer is simply a reusable part of code, and can be as simple as a fully connected neural network layer or as complex as a full inception network. We provide a library of layers which is well tested and whose implementation follow best practices. We have given our layers a consistent interface in order to ease the cognitive burden on users. In our framework, layers are implemented as free functions, taking \code{Tensor}s as input arguments (along with other parameters), and returning \code{Tensor}s. TensorFlow itself contains a large number of ops that behave in the same manner, so layers are a natural extension of TensorFlow and should feel natural to users of TensorFlow. Because layers accept and produce regular \code{Tensor}s, layers and regular TensorFlow ops can be mixed without requiring special care.

We implement layer functions with best practices in mind: layers are generally wrapped in a \code{variable\_scope}. This ensures that they are properly grouped in the TensorBoard visualization tool, which is essential when inspecting large models. All variables that are created as part of a layer are obtained using \code{get\_variable}, which ensures that variables can be reused or shared in different parts of the model. All layers assume that the first dimension of input tensors is the batch dimension, and accept variable batch size input. This allows changing the batch size as a hyperparameter during tuning, and it ensures that the model can be reused for inference, where inputs don't necessarily arrive in batches.

As an example, let's create a simple convolutional net to classify an image. The network comprises three convolutional and three pooling layers, as well as a final fully connected layer. We have set sensible defaults on many arguments, so the invocations are compact unless uncommon behavior is desired:

\begin{python}
# Input images as a 4D tensor (batch, width,
# height, and channels)
net = inputs
# instantiate 3 convolutional layers with pooling
for _ in range(3):
  net = layers.conv2d(net,
                      filters=4,
                      kernel_size=3,
                      activation=relu)
  net = layers.max_pooling2d(net,
                             pool_size=2,
                             strides=1)
logits = layers.dense(net, units=num_classes)
\end{python}

We separate out some classes of layers that share a more restricted interface. \textit{Losses} are functions which take an input, a label, and a weight, and return a scalar loss. These functions, such as \code{l1\_loss} or \code{l2\_loss} are used to produce a loss for optimization.

Metrics are another special class of layers commonly used in evaluation: they take again a label, a prediction, and optionally a weight, and compute a metric such as log-likelihood, accuracy, or a simple mean squared error. While superficially similar to losses, they support aggregating a metric across many minibatches, an important feature whenever the evaluation dataset does not fit into memory. Metrics return two \code{Tensor}s: \code{update\_op}, which should be run for each minibatch, and a \code{value\_op} which computes the final metric value. The \code{update\_op} does not return a value, and only updates internal variables, aggregating the new information contained in the input minibatch. The \code{value\_op} uses only the internal state to compute a metric value and returns it. The \code{Estimator}'s evaluation functionality relies on this usage pattern (see below). Properly implementing metrics is nontrivial, and our experience shows that metrics that are naively implemented from scratch lead to problems when using large datasets (using TensorFlow queues in evaluation requires extra finesse to avoid losing examples to logging or TensorBoard summary writing).

\subsection{Estimator}\label{estimator}
At the heart of our framework is \code{Estimator}, a class that both provides an interface for downstream infrastructure, as well as a convenient harness for developers. The interface for users of \code{Estimator} is loosely modeled after Scikit-learn and consists of only four methods: \code{train} trains the model, given training data. \code{evaluate} computes evaluation metrics over test data, \code{predict} performs inference on new data given a trained model, and finally, \code{export\_savedmodel} exports a SavedModel, a serialization format which allows the model to be used in TensorFlow Serving, a prebuilt production server for TensorFlow models~\cite{TensorFlowServingBlogPost}.

The user configures an \code{Estimator} by passing a callback, the \code{model\_fn}, to the constructor. When one of its methods is called, \code{Estimator} creates a TensorFlow graph, sets up the input pipeline specified by the user in the arguments to the method (see \refsec{inputs}), and then calls the \code{model\_fn} with appropriate arguments to generate the graph representing the model. The \code{Estimator} class itself contains the necessary code to run a training or evaluation loop, to predict using a trained model, or to export a prediction model for use in production.

\code{Estimator} hides some TensorFlow concepts, such as \code{Graph} and \code{Session}, from the user. The \code{Estimator} constructor also receives a configuration object called \code{RunConfig} which communicates everything that this \code{Estimator} needs to know about the environment in which the model will be run: how many workers are available, how often to save intermediate checkpoints, etc.

To ensure encapsulation, \code{Estimator} creates a new graph, and possibly restores from checkpoint, every time a method is called. Rebuilding the graph is expensive, and it could be cached to make it more economical to run, say, \code{evaluate} or \code{predict} in a loop. However, we found it very useful to explicitly recreate the graph, trading off performance for clarity. Even if we did not rebuild the graph, writing such loops is highly suboptimal in terms of performance. Making this cost very visible discourages users from accidentally writing badly performing code.

\begin{figure}[htbp]
  \includegraphics[width=0.5\textwidth]{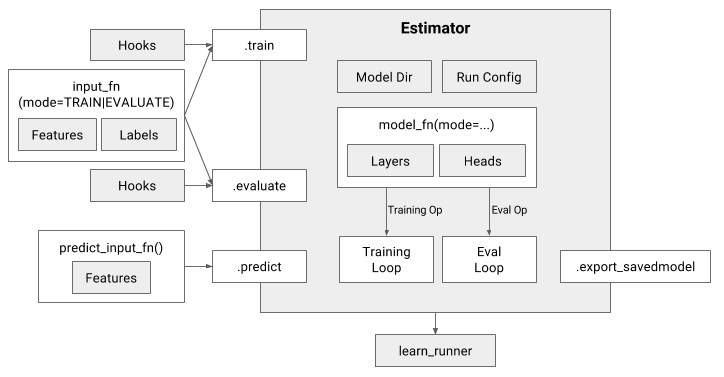}
  \caption{Simplified overview of the Estimator interface.}
  \label{fig:estimator-overview}
\end{figure}

A schematic of \code{Estimator} can be found in Figure~\ref{fig:estimator-overview}. Below, we first describe how to provide inputs to the \code{train}, \code{evaluate}, and \code{predict} methods using input functions. Then we discuss model specification with \code{model\_fn}, followed by how to specify outputs within the \code{model\_fn} using \code{Heads}.

\vspace{0.8em}
\noindent\textbf{Specifying inputs with input\_fn.}\label{inputs}
The methods \code{train}, \code{evaluate}, and \code{predict} all take an input function, which is expected to produce two dictionaries: one containing \code{Tensor}s with inputs (features), and one containing \code{Tensor}s with labels. Whenever a method of \code{Estimator} is called, a new graph is created, the \code{input\_fn} passed as an argument to the method call is called to produce the input pipeline of the \code{Estimator}, and then the \code{model\_fn} is called with the appropriate mode argument to build the actual model graph.
Decoupling the core model from input processing allows users to easily swap datasets. If used in larger infrastructure, being able to control the inputs completely is very valuable to downstream frameworks. A typical \code{input\_fn} has the following form:

\begin{python}
def my_input_fn(file_pattern):
    feature_dict = learn.io.read_batch_features(
        # path to data in tf.Example format
        file_pattern=file_pattern,
        batch_size=BATCH_SIZE,
        # whether sparse or dense ...
        features=FEATURE_SPEC,
        # such as TFRecordReader
        reader=READER,
        ...)

estimator.train(input_fn=lambda:
  my_input_fn(TRAINING_FILES), ...)
estimator.evaluate(input_fn=lambda:
  my_input_fn(EVAL_FILES), ...)
\end{python}

\vspace{0.8em}
\noindent\textbf{Specifying the model with model\_fn.}
We chose to configure \code{Estimator} with a single callback, the \code{model\_fn}, which returns ops for training, evaluation, or prediction, depending on which graph is being requested (which method of \code{Estimator} is being called). For example, if the \code{train} method is called, \code{model\_fn} will be called with an argument \code{mode=TRAIN}, which the user can then use to build a custom graph in the knowledge that it is going to be used for training.

Conceptually, three entirely different graphs can be built, and different information is returned, depending on the mode parameter representing the called method. Nevertheless, we found it useful to require only a single function for configuration. One of the main sources of error in production systems is training/serving skew. One type of training/serving skew happens when a different model is trained than is later served in production. Of course, models are routinely trained slightly differently than they are served. For instance, dropout and batch normalization layers are only active during training. However, it is easy to make mistakes if one has to rewrite the whole model three times. Therefore we chose to require a single function, effectively encouraging the model developer to write the model only once. For complex models, appropriate Python conditionals can be used to ensure that legitimate differences are explicitly represented in the model. A typical \code{model\_fn} for a simple model may look like this:
\begin{python}
def model_fn(features, target, mode, params):
  predictions = tf.stack(tf.fully_connected,
                         [50, 50, 1])
  loss = tf.losses.mean_squared_error(target,
                                      predictions)
  train_op = tf.train.create_train_op(
    loss, tf.train.get_global_step(),
    params['learning_rate'], params['optimizer'])
  return EstimatorSpec(mode=mode,
                       predictions=predictions,
                       loss=loss,
                       train_op=train_op)
\end{python}

\vspace{0.8em}
\noindent\textbf{Specifying outputs with Heads.}
The \code{Head} API is an abstraction for the part of the model behind the last hidden layer. The key goals of the design are to simplify writing \code{model\_fn}, to be compatible with a wide range of models, and to simplify supporting multiple heads.  A \code{Head} knows how to compute loss, relevant evaluation metrics, predictions and metadata about the predictions that other systems (like serving, model validation) can use. To support different types of models (e.g., DNN, linear, Wide \& Deep \cite{Cheng:2016:WDL:2988450.2988454}, gradient boosted trees, etc.), \code{Head} takes logits and labels as input and generates \code{Tensor}s for loss, metrics, and predictions. \code{Head}s can also take the activation of the last hidden layer as input to support DNN with large number of classes where we want to avoid computing the full logit \code{Tensor}. A typical \code{model\_fn} for a simple single objective model may look like this:
\begin{python}
def model_fn(features, target, mode, params):
  last_layer = tf.stack(tf.fully_connected,
                        [50, 50])
  head = tf.multi_class_head(n_classes=10)
  return head.create_estimator_spec(
      features, mode, last_layer,
      label=target,
      train_op_fn=lambda loss:
        my_optimizer.minimize(
            loss, tf.train.get_global_step())
\end{python}

The abstraction is designed in a way that combining multiple \code{Head}s for multi objective learning is as simple as creating a special type of \code{Head} with a list of other heads. Model functions can take \code{Head} as a parameter while remaining agnostic to what kind of \code{Head} they are using. A typical \code{model\_fn} for a simple model with two multi class objectives can look like this:
\begin{python}
def model_fn(features, target, mode, params):
  last_layer = tf.stack(tf.fully_connected,
                        [50, 50])
  head1 = tf.multi_class_head(n_classes=2,
            label_name='y', head_name='h1')
  head2 = tf.multi_class_head(n_classes=10,
            label_name='z', head_name='h2')
  head = tf.multi_head([head1, head2])
  return head.create_model_fn_ops(features,
       features, mode, last_layer,
      label=target,
      train_op_fn=lambda loss:
        my_optimizer.minimize(
            loss, tf.train.get_global_step())
\end{python}

\vspace{0.8em}
\noindent\textbf{Executing computations.}
Once the graph is built, the \code{Estimator} then initializes a \code{Session}, prepares it appropriately, and runs the training loop, evaluation loop, or iterates over the inputs to produce predictions.

Most machine learning algorithms are iterative nonlinear optimizations, and therefore have a particularly simple algorithmic form: a single loop which runs the same computation over and over again, with different input data in each iteration. When used during training, this is called the training loop. In evaluation using mini-batches, much the same structure is used, except that variables are not updated, and typically, more metrics than just the loss are computed.

An idealized training loop implemented in TensorFlow is simple: start a \code{Session}, then run a training op in a loop. However, we have to at least initialize variables and special data structures like tables which are used in embeddings. Queue runners (implemented as Python threads) have to be started, and should be stopped at the end to ensure a clean exit. Summaries (which provide data to the TensorBoard visualization tool) have to be computed and written to file. The real challenge begins when distributed training is taken into account. While TensorFlow takes care of distribution of the computation and communication between workers, it requires many coordinated steps before a model can be successfully trained. The distributed computation introduces a number of opportunities for users to make mistakes: certain variables must be initialized on all workers, most only on one. The model state should be saved periodically to ensure that the computation can recover when workers go down, and needs to be recovered safely when they restart. End-of-input signals have to be handled gracefully.

Because the training loop is so ubiquitous, a good implementation removes a lot of duplicated user code. Because it is simple only in theory, we can remove a source of error and frustration for users. Therefore, \code{Estimator} implements and controls the training loop. It automatically assigns \code{Variables} to parameter servers to simplify distributed computation, and it gives the user only limited access to the underlying TensorFlow primitives. Users must specify the graph, and the op(s) to run in each iteration, and they may override the device placement.

\vspace{0.8em}
\noindent\textbf{Code injection using Hooks.}
\code{Hook}s make it impossible to implement advanced optimization techniques that break the simple loop abstraction in a safe manner. They are also useful for custom processing that has to happen alongside the main loop, for recordkeeping, debugging, monitoring or reporting. \code{Hook}s let users define custom behaviour at \code{Session} creation, before and after each iteration, and at the end of training. They also let users add ops other than those specified by the \code{model\_fn} to be run within the same \code{Session.run} call. For example, a user who wants to train not for a given number of steps, but a given amount of wall time, could implement a \code{Hook} as follows:

\begin{python}
class TimeBasedStopHook(tf.train.SessionRunHook):
  def begin(self):
    self.started_at = time.time()
  def after_run(self, run_context, run_values):
    if time.time() - self.started_at >= TRAIN_TIME:
      run_context.request_stop()
\end{python}

\code{Hook}s are activated by passing them to the \code{train} call. When the \code{Hook} shown above is passed to \code{train}, the model training will end after the set time. Much of the functionality that \code{Estimator} provides (for instance, summaries, step counting, and checkpointing) is internally implemented using such \code{Hook}s.

\subsection{Canned Estimators}\label{cannedestimators}

There are many model architectures commonly used by researchers and practitioners. We decided to provide those architectures as canned \code{Estimator}s so that users don't need to rewrite the same models again and again. Canned \code{Estimator}s are a good example of how to use \code{Estimator} itself. They are direct subclasses of \code{Estimator} that only override their constructors. As such, users of canned \code{Estimator}s would only need to know how to use an \code{Estimator}, and how to configure the canned \code{Estimator}. This means that canned \code{Estimator}s are mainly restricted to define a canned \code{model\_fn}. There are two main reasons behind this restrictive design. First, we are expecting an increasing number of canned \code{Estimator}s to be implemented. To minimize the cognitive load on users, all these canned \code{Estimator}s should behave identically. Second, this restriction makes the canned \code{Estimator} developer a user of \code{Estimator}. This leads to an implicit comprehensive flexibility test of our API.

Neural networks rely on operations which take dense \code{Tensor}s and output dense \code{Tensor}s. Many machine learning problems have sparse features such as query keywords, product id, url, video id, etc. For models with many inputs, specifying how these features are attached to the model often consumes a large fraction of the total setup time. Based on our experience, one of the most error prone parts of building a model is converting these features into a single dense \code{Tensor}.

We offer the \code{FeatureColumn} abstraction to simplify input ingestion. \code{FeatureColumn}s are a declarative way of specifying inputs. Canned \code{Estimator}s take \code{FeatureColumn}s as a constructor argument and handle the conversion of sparse or dense features of all types to a dense \code{Tensor} usable by the core model. As an example, the following code shows a canned \code{Estimator} implementation for the Wide \& Deep architecture \cite{Cheng:2016:WDL:2988450.2988454}. The deep part of the model uses embeddings while the linear part uses the crosses of base features.

\vspace{0.8em}
\begin{python}
# Define wide model features and crosses.
query_x_docid = crossed_column(
    ["query", "docid"], num_buckets)
wide_cols = [query_x_docid, ...]

# Define deep model features and embeddings.
query = categorical_column_with_hash_bucket(
    "query", num_buckets)
docid = categorical_column_with_hash_bucket(
    "docid", num_buckets)
query_emb = embedding_column(query, dimension=32)
docid_emb = embedding_column(docid, dimension=32)
deep_cols = [query_emb, docid_emb, ...]
# Define model structure and start training.
estimator = DNNLinearCombinedClassifier(
    wide_cols, deep_cols,
    dnn_hidden_units=[500, 200, 100])
estimator.train(input_fn, ...)
\end{python}

\section{Distributed Execution}\label{distributed}
\begin{figure}[htbp]
  \includegraphics[width=0.5\textwidth]{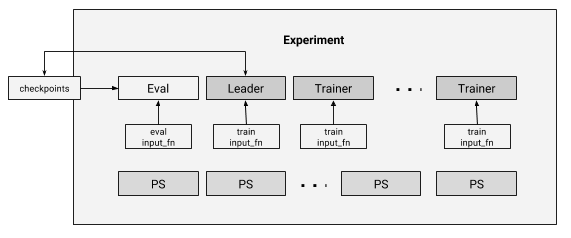}
  \caption{Simplified overview of the Experiment interface.}
  \label{fig:experiment-overview}
\end{figure}

With the built-in functionalities and utilities mentioned above, \code{Estimator}s are ready for training, evaluating and exporting the model on a single machine. For production usages and models with large amounts of training data, utilities for distributed execution are also provided together with Estimators, which takes the advantage of TensorFlow's distributed training support. The core of distributed execution support is the \code{Experiment} class, which groups the \code{Estimator} with two input functions for training and evaluation. The architecture is summarized in Figure~\ref{fig:experiment-overview}.

In each TensorFlow cluster, there are several parameter servers and several worker tasks. Most workers are handing the training process, which basically calls the \code{Estimator} \code{train} method with the training \code{input{\_}fn}. One of the workers is designated leader and is responsible for managing checkpoints and other maintenance work. Currently, the primary mode of replica training in TensorFlow Estimators is between-graph replication and asynchronous training. However, it could be easily extended to support other replicated training settings. With this architecture, gradient descent training can be executed in parallel.

\begin{figure}[htbp]
  \includegraphics[width=0.5\textwidth]{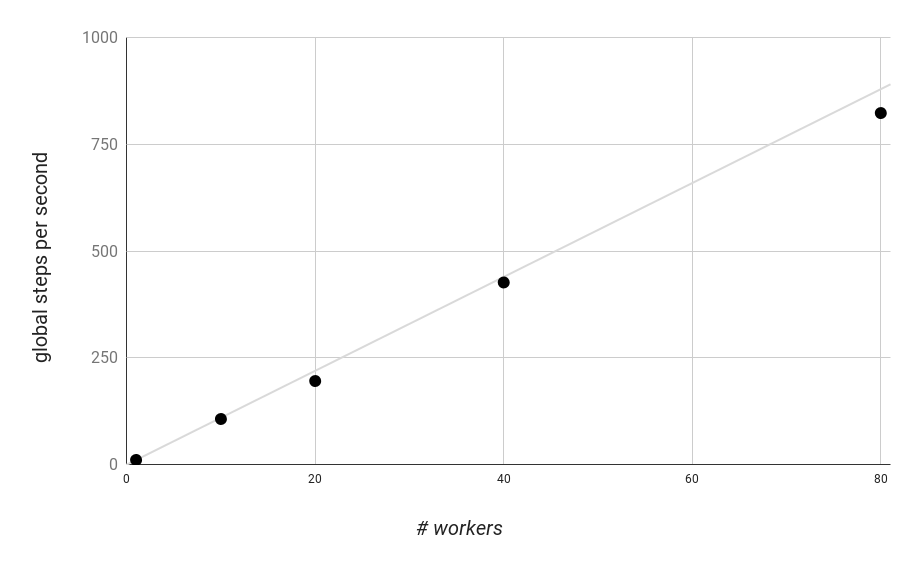}
  \caption{Measuring scaling of DNN model training implemented with TensorFlow Estimators, varying the number of workers. Shown are measurements as well as the theoretical perfect linear scaling.}
  \label{fig:scale}
\end{figure}

We have evaluated scaling of TensorFlow Estimators by running different numbers of workers with fixed numbers of parameter servers. We trained a DNN model on a large internal recommendation dataset (100s of billions of examples) for 48 hours and present average number of training steps per second. Figure~\ref{fig:scale} shows that we achieve almost linear scaling of global steps per second with the number of workers.

There is a special worker handling the evaluation process for the \code{Experiment} to evaluate the performance and export the model. It runs in a continuous loop and calls the \code{Estimator} evaluate method with the evaluation \code{input{\_}fn}. In order to avoid race conditions and inconsistent model parameter states, the evaluation process always begins with loading the latest checkpoint and calculates the evaluation metrics based on the model parameters from that checkpoint. As a simple extension, the \code{Experiment} also supports the evaluation with the training \code{input{\_}fn}, which is very useful to detect overfitting in deep learning in practice.

Furthermore, we also provide utilities, \code{RunConfig} and \code{runner}, to ease the way of using and configuring \code{Experiment} in a cluster for distributed training. \code{RunConfig} holds all the execution related configuration the \code{Experiment}/\code{Estimator} requires, including cluster specification, model output directory, checkpoints configuration, etc. In particular, \code{RunConfig} specifies the task type of the current task, which allows all tasks sharing the same binary but running a different mode, such as parameter server, training, or continual evaluation. The \code{runner} is simply a utility method to construct the \code{RunConfig}, e.g., by parsing the environment variable, and execute the \code{Experiment}/\code{Estimator} with that \code{RunConfig}. With this design,  \code{Experiment}/\code{Estimator} could be easily shared by various execution frameworks including end-to-end machine learning pipelines \cite{tfx} and even hyper-parameters tuning.

\section{Case Studies and Adoption}\label{case}
For machine learning practitioners within Google, this framework has dramatically reduced the time to launch a working model. Before TensorFlow Estimators, the typical model construction cycle involved writing custom TensorFlow code to ingest and represent features (sparse features were especially tricky), construction of the model layers itself, establishing training and validation loops, productionizing the system to run on distributed training clusters, adding evaluation metrics, debugging training NaNs, and debugging poor model quality.

TensorFlow Estimators simplify or automate all but the debugging steps. Estimators give the practitioner confidence that, when debugging NaNs or poor quality, these problems arise either from their choice of hyperparameters or their choice of features --- but not a bug in the wiring of the model itself.

When TensorFlow Estimators became available, several TensorFlow models under development greatly benefited from transitioning to the framework. One multiclass classification model attained 37\% better model accuracy by switching from a custom model that performed multiple logistic regressions to a standard \code{Estimator} that properly used a softmax cross-entropy loss --- the switch also reduced lines of code required from 800 to 200. A different TensorFlow CTR model was stuck in the debugging phase for several weeks, but was transitioned to the framework within two days and achieved launchable offline metrics.

It is worth noting that using \code{Estimators} and the associated machinery also requires considerably less expertise than would be required to implement the equivalent functionality from scratch. Recently, a cohort of Google data scientists with limited Python experience and no TensorFlow experience were able to bootstrap real models in a two-day class setting.

\subsection{Experience in YouTube Watch Next}
Using TensorFlow Estimators, we have productionized and launched a deep model (\code{DNNClassifier}) in the Watch Next video recommender system of YouTube. Watch Next is a product recommending a ranked set of videos for a user to choose from after the user is done watching the current video. One unique aspect about our model is that the model is trained over multiple days, with the training data being continuously updated.

Our input features consist of both sparse categorical features and real-valued features. The sparse features are further transformed into embedding columns before being fed into the hidden layers. The \code{FeatureColumn} API greatly simplifies how we construct the input layer of our model. Additionally, the train-to-serve support of TensorFlow Estimators considerably reduced the engineering effort to productionize the Watch Next model. Furthermore, the \code{Estimator} framework made it easy to implement new \code{Estimator}s and experiment with new model architectures such as multiple-objective learning to accommodate specific product needs.

The initial version of the model pipeline was developed using low-level TensorFlow primitives prior to the release of Estimators. While debugging why the model quality failed to match our expectation, we discovered critical bugs related to how the network layers were constructed and how the input data were processed.

As an early adopter, Watch Next prompted the development of missing features such as shared embedding columns. Shared embedding columns allow multiple semantically similar features to share a common embedding space, with the benefit of transfer learning across features and smaller model size.

\subsection{Adoption within Google}
Software engineers at Google have a variety of choices for how to implement their machine learning models. Before we developed the higher-level framework in TensorFlow, engineers were effectively forced to implement one-off versions of the components in our framework.

\begin{figure}[htbp]
  \includegraphics[width=0.5\textwidth]{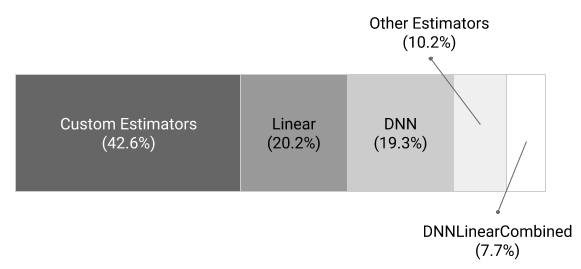}
  \caption{Current usage of Estimators at Google.}
  \label{fig:usage}
\end{figure}

An internal survey has shown that, since we introduced this framework and \code{Estimator}s less than a year ago, close to 1,000 \code{Estimator}s have been checked into the Google codebase and more than 120,000 experiments have been recorded (an experiment in this context is a complete training run; not all runs are recorded, so the true number is significantly higher). Of those, over half (57\%) use implementations of canned \code{Estimator}s (e.g., \code{LinearClassifier}, \code{DNNLinearCombinedRegressor}). There are now over 20 Estimator classes implementing various standard machine learning algorithms in the TensorFlow code base. Examples include \code{DynamicRnnEstimator} (implementing dynamically unrolled RNNs for classification or regression problems) and \code{TensorForestEstimator} (implementing random forests). Figure~\ref{fig:usage} shows the current distribution of \code{Estimator} usage. This framework allowed teams to build high-quality machine learning models within an average of one engineer-week, sometimes as fast as within 2 hours. 74\% of respondents say that development with this framework is faster than other machine learning APIs they used before. Most importantly, users note that they can focus their time on the machine learning problem as opposed to the implementation of underlying basics. Among existing users, quick ramp-up, ease of use, reuse of common code and readability of a commonly used framework are the most frequently mentioned benefits.

\Urlmuskip=0mu plus 1mu\relax
\bibliographystyle{ACM_Reference_Format}
\bibliography{main}


\begin{thebibliography}{00}


\ifx \showCODEN    \undefined \def \showCODEN     #1{\unskip}     \fi
\ifx \showDOI      \undefined \def \showDOI       #1{{\tt DOI:}\penalty0{#1}\ }
  \fi
\ifx \showISBNx    \undefined \def \showISBNx     #1{\unskip}     \fi
\ifx \showISBNxiii \undefined \def \showISBNxiii  #1{\unskip}     \fi
\ifx \showISSN     \undefined \def \showISSN      #1{\unskip}     \fi
\ifx \showLCCN     \undefined \def \showLCCN      #1{\unskip}     \fi
\ifx \shownote     \undefined \def \shownote      #1{#1}          \fi
\ifx \showarticletitle \undefined \def \showarticletitle #1{#1}   \fi
\ifx \showURL      \undefined \def \showURL       #1{#1}          \fi
\providecommand\bibfield[2]{#2}
\providecommand\bibinfo[2]{#2}
\providecommand\natexlab[1]{#1}
\providecommand\showeprint[2][]{arXiv:#2}

\bibitem[\protect\citeauthoryear{??}{Ten}{}]%
        {TensorFlowServingBlogPost}
\bibinfo{booktitle}{{\em Running your models in production with TensorFlow
  Serving}}.
\newblock
\newblock
\shownote{\url{https://research.googleblog.com/2016/02/running-your-models-in-production-with.html},
  accessed 2017-02-08.}


\bibitem[\protect\citeauthoryear{Abadi, Barham, Chen, Chen, Davis, Dean, Devin,
  Ghemawat, Irving, Isard, Kudlur, Levenberg, Monga, Moore, Murray, Steiner,
  Tucker, Vasudevan, Warden, Wicke, Yu, and Zheng}{Abadi et~al\mbox{.}}{2016}]%
        {tensorflow}
\bibfield{author}{\bibinfo{person}{Mart{\'{\i}}n Abadi}, \bibinfo{person}{Paul
  Barham}, \bibinfo{person}{Jianmin Chen}, \bibinfo{person}{Zhifeng Chen},
  \bibinfo{person}{Andy Davis}, \bibinfo{person}{Jeffrey Dean},
  \bibinfo{person}{Matthieu Devin}, \bibinfo{person}{Sanjay Ghemawat},
  \bibinfo{person}{Geoffrey Irving}, \bibinfo{person}{Michael Isard},
  \bibinfo{person}{Manjunath Kudlur}, \bibinfo{person}{Josh Levenberg},
  \bibinfo{person}{Rajat Monga}, \bibinfo{person}{Sherry Moore},
  \bibinfo{person}{Derek~Gordon Murray}, \bibinfo{person}{Benoit Steiner},
  \bibinfo{person}{Paul~A. Tucker}, \bibinfo{person}{Vijay Vasudevan},
  \bibinfo{person}{Pete Warden}, \bibinfo{person}{Martin Wicke},
  \bibinfo{person}{Yuan Yu}, {and} \bibinfo{person}{Xiaoqiang Zheng}.}
  \bibinfo{year}{2016}\natexlab{}.
\newblock \showarticletitle{TensorFlow: {A} System for Large-Scale Machine
  Learning}. In \bibinfo{booktitle}{{\em OSDI}}. \bibinfo{pages}{265--283}.
\newblock


\bibitem[\protect\citeauthoryear{Agarwal, Akchurin, Basoglu, Chen, Cyphers,
  Droppo, Eversole, Guenter, Hillebrand, Hoens, Huang, Huang, Ivanov, Kamenev,
  Kranen, Kuchaiev, Manousek, May, Mitra, Nano, Navarro, Orlov, Padmilac,
  Parthasarathi, Peng, Reznichenko, Seide, Seltzer, Slaney, Stolcke, Wang,
  Wang, Yao, Yu, Zhang, and Zweig}{Agarwal et~al\mbox{.}}{2014}]%
        {cntk}
\bibfield{author}{\bibinfo{person}{Amit Agarwal}, \bibinfo{person}{Eldar
  Akchurin}, \bibinfo{person}{Chris Basoglu}, \bibinfo{person}{Guoguo Chen},
  \bibinfo{person}{Scott Cyphers}, \bibinfo{person}{Jasha Droppo},
  \bibinfo{person}{Adam Eversole}, \bibinfo{person}{Brian Guenter},
  \bibinfo{person}{Mark Hillebrand}, \bibinfo{person}{Ryan Hoens},
  \bibinfo{person}{Xuedong Huang}, \bibinfo{person}{Zhiheng Huang},
  \bibinfo{person}{Vladimir Ivanov}, \bibinfo{person}{Alexey Kamenev},
  \bibinfo{person}{Philipp Kranen}, \bibinfo{person}{Oleksii Kuchaiev},
  \bibinfo{person}{Wolfgang Manousek}, \bibinfo{person}{Avner May},
  \bibinfo{person}{Bhaskar Mitra}, \bibinfo{person}{Olivier Nano},
  \bibinfo{person}{Gaizka Navarro}, \bibinfo{person}{Alexey Orlov},
  \bibinfo{person}{Marko Padmilac}, \bibinfo{person}{Hari Parthasarathi},
  \bibinfo{person}{Baolin Peng}, \bibinfo{person}{Alexey Reznichenko},
  \bibinfo{person}{Frank Seide}, \bibinfo{person}{Michael~L. Seltzer},
  \bibinfo{person}{Malcolm Slaney}, \bibinfo{person}{Andreas Stolcke},
  \bibinfo{person}{Yongqiang Wang}, \bibinfo{person}{Huaming Wang},
  \bibinfo{person}{Kaisheng Yao}, \bibinfo{person}{Dong Yu},
  \bibinfo{person}{Yu Zhang}, {and} \bibinfo{person}{Geoffrey Zweig}.}
  \bibinfo{year}{2014}\natexlab{}.
\newblock \bibinfo{booktitle}{{\em An Introduction to Computational Networks
  and the Computational Network Toolkit}}.
\newblock \bibinfo{type}{{T}echnical {R}eport} MSR-TR-2014-112.
\newblock
\showURL{%
\url{http://research.microsoft.com/apps/pubs/default.aspx?id=226641}}


\bibitem[\protect\citeauthoryear{Al-Rfou, Alain, Almahairi, Angermueller,
  Bahdanau, Ballas, Bastien, Bayer, Belikov, Belopolsky, Bengio, Bergeron,
  Bergstra, Bisson, {Bleecher Snyder}, Bouchard, Boulanger-Lewandowski,
  Bouthillier, de~Br\'ebisson, Breuleux, Carrier, Cho, Chorowski, Christiano,
  Cooijmans, C\^ot\'e, C\^ot\'e, Courville, Dauphin, Delalleau, Demouth,
  Desjardins, Dieleman, Dinh, Ducoffe, Dumoulin, {Ebrahimi Kahou}, Erhan, Fan,
  Firat, Germain, Glorot, Goodfellow, Graham, Gulcehre, Hamel, Harlouchet,
  Heng, Hidasi, Honari, Jain, Jean, Jia, Korobov, Kulkarni, Lamb, Lamblin,
  Larsen, Laurent, Lee, Lefrancois, Lemieux, L\'eonard, Lin, Livezey, Lorenz,
  Lowin, Ma, Manzagol, Mastropietro, McGibbon, Memisevic, van Merri\"enboer,
  Michalski, Mirza, Orlandi, Pal, Pascanu, Pezeshki, Raffel, Renshaw, Rocklin,
  Romero, Roth, Sadowski, Salvatier, Savard, Schl\"uter, Schulman, Schwartz,
  Serban, Serdyuk, Shabanian, Simon, Spieckermann, Subramanyam, Sygnowski,
  Tanguay, van Tulder, Turian, Urban, Vincent, Visin, de~Vries, Warde-Farley,
  Webb, Willson, Xu, Xue, Yao, Zhang, and Zhang}{Al-Rfou et~al\mbox{.}}{2016}]%
        {theano}
\bibfield{author}{\bibinfo{person}{Rami Al-Rfou}, \bibinfo{person}{Guillaume
  Alain}, \bibinfo{person}{Amjad Almahairi}, \bibinfo{person}{Christof
  Angermueller}, \bibinfo{person}{Dzmitry Bahdanau}, \bibinfo{person}{Nicolas
  Ballas}, \bibinfo{person}{Fr\'ed\'eric Bastien}, \bibinfo{person}{Justin
  Bayer}, \bibinfo{person}{Anatoly Belikov}, \bibinfo{person}{Alexander
  Belopolsky}, \bibinfo{person}{Yoshua Bengio}, \bibinfo{person}{Arnaud
  Bergeron}, \bibinfo{person}{James Bergstra}, \bibinfo{person}{Valentin
  Bisson}, \bibinfo{person}{Josh {Bleecher Snyder}}, \bibinfo{person}{Nicolas
  Bouchard}, \bibinfo{person}{Nicolas Boulanger-Lewandowski},
  \bibinfo{person}{Xavier Bouthillier}, \bibinfo{person}{Alexandre de
  Br\'ebisson}, \bibinfo{person}{Olivier Breuleux}, \bibinfo{person}{Pierre-Luc
  Carrier}, \bibinfo{person}{Kyunghyun Cho}, \bibinfo{person}{Jan Chorowski},
  \bibinfo{person}{Paul Christiano}, \bibinfo{person}{Tim Cooijmans},
  \bibinfo{person}{Marc-Alexandre C\^ot\'e}, \bibinfo{person}{Myriam C\^ot\'e},
  \bibinfo{person}{Aaron Courville}, \bibinfo{person}{Yann~N. Dauphin},
  \bibinfo{person}{Olivier Delalleau}, \bibinfo{person}{Julien Demouth},
  \bibinfo{person}{Guillaume Desjardins}, \bibinfo{person}{Sander Dieleman},
  \bibinfo{person}{Laurent Dinh}, \bibinfo{person}{M\'elanie Ducoffe},
  \bibinfo{person}{Vincent Dumoulin}, \bibinfo{person}{Samira {Ebrahimi
  Kahou}}, \bibinfo{person}{Dumitru Erhan}, \bibinfo{person}{Ziye Fan},
  \bibinfo{person}{Orhan Firat}, \bibinfo{person}{Mathieu Germain},
  \bibinfo{person}{Xavier Glorot}, \bibinfo{person}{Ian Goodfellow},
  \bibinfo{person}{Matt Graham}, \bibinfo{person}{Caglar Gulcehre},
  \bibinfo{person}{Philippe Hamel}, \bibinfo{person}{Iban Harlouchet},
  \bibinfo{person}{Jean-Philippe Heng}, \bibinfo{person}{Bal\'azs Hidasi},
  \bibinfo{person}{Sina Honari}, \bibinfo{person}{Arjun Jain},
  \bibinfo{person}{S\'ebastien Jean}, \bibinfo{person}{Kai Jia},
  \bibinfo{person}{Mikhail Korobov}, \bibinfo{person}{Vivek Kulkarni},
  \bibinfo{person}{Alex Lamb}, \bibinfo{person}{Pascal Lamblin},
  \bibinfo{person}{Eric Larsen}, \bibinfo{person}{C\'esar Laurent},
  \bibinfo{person}{Sean Lee}, \bibinfo{person}{Simon Lefrancois},
  \bibinfo{person}{Simon Lemieux}, \bibinfo{person}{Nicholas L\'eonard},
  \bibinfo{person}{Zhouhan Lin}, \bibinfo{person}{Jesse~A. Livezey},
  \bibinfo{person}{Cory Lorenz}, \bibinfo{person}{Jeremiah Lowin},
  \bibinfo{person}{Qianli Ma}, \bibinfo{person}{Pierre-Antoine Manzagol},
  \bibinfo{person}{Olivier Mastropietro}, \bibinfo{person}{Robert~T. McGibbon},
  \bibinfo{person}{Roland Memisevic}, \bibinfo{person}{Bart van Merri\"enboer},
  \bibinfo{person}{Vincent Michalski}, \bibinfo{person}{Mehdi Mirza},
  \bibinfo{person}{Alberto Orlandi}, \bibinfo{person}{Christopher Pal},
  \bibinfo{person}{Razvan Pascanu}, \bibinfo{person}{Mohammad Pezeshki},
  \bibinfo{person}{Colin Raffel}, \bibinfo{person}{Daniel Renshaw},
  \bibinfo{person}{Matthew Rocklin}, \bibinfo{person}{Adriana Romero},
  \bibinfo{person}{Markus Roth}, \bibinfo{person}{Peter Sadowski},
  \bibinfo{person}{John Salvatier}, \bibinfo{person}{Fran\c{c}ois Savard},
  \bibinfo{person}{Jan Schl\"uter}, \bibinfo{person}{John Schulman},
  \bibinfo{person}{Gabriel Schwartz}, \bibinfo{person}{Iulian~Vlad Serban},
  \bibinfo{person}{Dmitriy Serdyuk}, \bibinfo{person}{Samira Shabanian},
  \bibinfo{person}{\'Etienne Simon}, \bibinfo{person}{Sigurd Spieckermann},
  \bibinfo{person}{S.~Ramana Subramanyam}, \bibinfo{person}{Jakub Sygnowski},
  \bibinfo{person}{J\'er\'emie Tanguay}, \bibinfo{person}{Gijs van Tulder},
  \bibinfo{person}{Joseph Turian}, \bibinfo{person}{Sebastian Urban},
  \bibinfo{person}{Pascal Vincent}, \bibinfo{person}{Francesco Visin},
  \bibinfo{person}{Harm de Vries}, \bibinfo{person}{David Warde-Farley},
  \bibinfo{person}{Dustin~J. Webb}, \bibinfo{person}{Matthew Willson},
  \bibinfo{person}{Kelvin Xu}, \bibinfo{person}{Lijun Xue}, \bibinfo{person}{Li
  Yao}, \bibinfo{person}{Saizheng Zhang}, {and} \bibinfo{person}{Ying Zhang}.}
  \bibinfo{year}{2016}\natexlab{}.
\newblock \showarticletitle{{Theano: A {Python} framework for fast computation
  of mathematical expressions}}.
\newblock \bibinfo{journal}{{\em arXiv e-prints\/}}
  \bibinfo{volume}{abs/1605.02688} (\bibinfo{date}{May} \bibinfo{year}{2016}).
\newblock
\showURL{%
\url{http://arxiv.org/abs/1605.02688}}


\bibitem[\protect\citeauthoryear{Amazon}{Amazon}{2016}]%
        {dsstne}
\bibfield{author}{\bibinfo{person}{Amazon}.} \bibinfo{year}{2016}\natexlab{}.
\newblock \bibinfo{title}{Dsstne}.
\newblock \bibinfo{howpublished}{https://github.com/amznlabs/amazon-dsstne}.
  (\bibinfo{year}{2016}).
\newblock


\bibitem[\protect\citeauthoryear{Baylor, Breck, Cheng, Fiedel, Foo, Haque,
  Haykal, Ispir, Jain, Koc, Koo, Lew, Mewald, Modi, Polyzotis, Ramesh, Roy,
  Whang, Wicke, Wilkiewicz, Zhang, and Zinkevich}{Baylor et~al\mbox{.}}{2017}]%
        {tfx}
\bibfield{author}{\bibinfo{person}{Denis Baylor}, \bibinfo{person}{Eric Breck},
  \bibinfo{person}{Heng-Tze Cheng}, \bibinfo{person}{Noah Fiedel},
  \bibinfo{person}{Chuan~Yu Foo}, \bibinfo{person}{Zakaria Haque},
  \bibinfo{person}{Salem Haykal}, \bibinfo{person}{Mustafa Ispir},
  \bibinfo{person}{Vihan Jain}, \bibinfo{person}{Levent Koc},
  \bibinfo{person}{Chiu~Yuen Koo}, \bibinfo{person}{Lukasz Lew},
  \bibinfo{person}{Clemens Mewald}, \bibinfo{person}{Akshay~Naresh Modi},
  \bibinfo{person}{Neoklis Polyzotis}, \bibinfo{person}{Sukriti Ramesh},
  \bibinfo{person}{Sudip Roy}, \bibinfo{person}{Steven~Euijong Whang},
  \bibinfo{person}{Martin Wicke}, \bibinfo{person}{Jarek Wilkiewicz},
  \bibinfo{person}{Xin Zhang}, {and} \bibinfo{person}{Martin Zinkevich}.}
  \bibinfo{year}{2017}\natexlab{}.
\newblock \bibinfo{title}{The Anatomy of a Production-Scale
  Continuously-Training ML Platform}.
\newblock \bibinfo{howpublished}{KDD [under review]}.   (\bibinfo{year}{2017}).
\newblock


\bibitem[\protect\citeauthoryear{Chen and Guestrin}{Chen and Guestrin}{2016}]%
        {xgboost}
\bibfield{author}{\bibinfo{person}{Tianqi Chen} {and} \bibinfo{person}{Carlos
  Guestrin}.} \bibinfo{year}{2016}\natexlab{}.
\newblock \showarticletitle{XGBoost: {A} Scalable Tree Boosting System}.
\newblock \bibinfo{journal}{{\em CoRR\/}}  \bibinfo{volume}{abs/1603.02754}
  (\bibinfo{year}{2016}).
\newblock
\showURL{%
\url{http://arxiv.org/abs/1603.02754}}


\bibitem[\protect\citeauthoryear{Chen, Li, Li, Lin, Wang, Wang, Xiao, Xu,
  Zhang, and Zhang}{Chen et~al\mbox{.}}{2015}]%
        {mxnet}
\bibfield{author}{\bibinfo{person}{Tianqi Chen}, \bibinfo{person}{Mu Li},
  \bibinfo{person}{Yutian Li}, \bibinfo{person}{Min Lin},
  \bibinfo{person}{Naiyan Wang}, \bibinfo{person}{Minjie Wang},
  \bibinfo{person}{Tianjun Xiao}, \bibinfo{person}{Bing Xu},
  \bibinfo{person}{Chiyuan Zhang}, {and} \bibinfo{person}{Zheng Zhang}.}
  \bibinfo{year}{2015}\natexlab{}.
\newblock \showarticletitle{MXNet: {A} Flexible and Efficient Machine Learning
  Library for Heterogeneous Distributed Systems}.
\newblock \bibinfo{journal}{{\em CoRR\/}}  \bibinfo{volume}{abs/1512.01274}
  (\bibinfo{year}{2015}).
\newblock
\showURL{%
\url{http://arxiv.org/abs/1512.01274}}


\bibitem[\protect\citeauthoryear{Cheng, Koc, Harmsen, Shaked, Chandra, Aradhye,
  Anderson, Corrado, Chai, Ispir, Anil, Haque, Hong, Jain, Liu, and Shah}{Cheng
  et~al\mbox{.}}{2016}]%
        {Cheng:2016:WDL:2988450.2988454}
\bibfield{author}{\bibinfo{person}{Heng-Tze Cheng}, \bibinfo{person}{Levent
  Koc}, \bibinfo{person}{Jeremiah Harmsen}, \bibinfo{person}{Tal Shaked},
  \bibinfo{person}{Tushar Chandra}, \bibinfo{person}{Hrishi Aradhye},
  \bibinfo{person}{Glen Anderson}, \bibinfo{person}{Greg Corrado},
  \bibinfo{person}{Wei Chai}, \bibinfo{person}{Mustafa Ispir},
  \bibinfo{person}{Rohan Anil}, \bibinfo{person}{Zakaria Haque},
  \bibinfo{person}{Lichan Hong}, \bibinfo{person}{Vihan Jain},
  \bibinfo{person}{Xiaobing Liu}, {and} \bibinfo{person}{Hemal Shah}.}
  \bibinfo{year}{2016}\natexlab{}.
\newblock \showarticletitle{Wide \& Deep Learning for Recommender Systems}. In
  \bibinfo{booktitle}{{\em DLRS}}. \bibinfo{pages}{7--10}.
\newblock


\bibitem[\protect\citeauthoryear{Chollet}{Chollet}{2015}]%
        {keras}
\bibfield{author}{\bibinfo{person}{Fran{\c c}ois Chollet}.}
  \bibinfo{year}{2015}\natexlab{}.
\newblock \bibinfo{title}{keras}.
\newblock \bibinfo{howpublished}{\url{https://github.com/fchollet/keras}}.
  (\bibinfo{year}{2015}).
\newblock


\bibitem[\protect\citeauthoryear{Collobert, Bengio, and Marithoz}{Collobert
  et~al\mbox{.}}{2002}]%
        {torch}
\bibfield{author}{\bibinfo{person}{Ronan Collobert}, \bibinfo{person}{Samy
  Bengio}, {and} \bibinfo{person}{Johnny Marithoz}.}
  \bibinfo{year}{2002}\natexlab{}.
\newblock \bibinfo{title}{Torch: A Modular Machine Learning Software Library}.
\newblock   (\bibinfo{year}{2002}).
\newblock


\bibitem[\protect\citeauthoryear{community}{community}{2012}]%
        {numpy}
\bibfield{author}{\bibinfo{person}{The~Scipy community}.}
  \bibinfo{year}{2012}\natexlab{}.
\newblock \bibinfo{booktitle}{{\em {NumPy Reference Guide}}}.
\newblock \bibinfo{publisher}{SciPy.org}.
\newblock
\showURL{%
\url{http://docs.scipy.org/doc/numpy/reference/}}


\bibitem[\protect\citeauthoryear{Dean, Corrado, Monga, Chen, Devin, Le, Mao,
  Ranzato, Senior, Tucker, Yang, and Ng}{Dean et~al\mbox{.}}{2012}]%
        {distbelief}
\bibfield{author}{\bibinfo{person}{Jeffrey Dean}, \bibinfo{person}{Greg~S.
  Corrado}, \bibinfo{person}{Rajat Monga}, \bibinfo{person}{Kai Chen},
  \bibinfo{person}{Matthieu Devin}, \bibinfo{person}{Quoc~V. Le},
  \bibinfo{person}{Mark~Z. Mao}, \bibinfo{person}{Marc'Aurelio Ranzato},
  \bibinfo{person}{Andrew Senior}, \bibinfo{person}{Paul Tucker},
  \bibinfo{person}{Ke Yang}, {and} \bibinfo{person}{Andrew~Y. Ng}.}
  \bibinfo{year}{2012}\natexlab{}.
\newblock \showarticletitle{Large Scale Distributed Deep Networks}. In
  \bibinfo{booktitle}{{\em Proceedings of the 25th International Conference on
  Neural Information Processing Systems}} {\em (\bibinfo{series}{NIPS'12})}.
  \bibinfo{publisher}{Curran Associates Inc.}, \bibinfo{address}{USA},
  \bibinfo{pages}{1223--1231}.
\newblock
\showURL{%
\url{http://dl.acm.org/citation.cfm?id=2999134.2999271}}


\bibitem[\protect\citeauthoryear{{Deeplearning4j Development
  Team.}}{{Deeplearning4j Development Team.}}{2016}]%
        {dl4j}
\bibfield{author}{\bibinfo{person}{{Deeplearning4j Development Team.}}}
  \bibinfo{year}{2016}\natexlab{}.
\newblock \bibinfo{title}{Deeplearning4j: Open-source distributed deep learning
  for the JVM, Apache Software Foundation License 2.0.}
\newblock \bibinfo{howpublished}{http://deeplearning4j.org}.
  (\bibinfo{year}{2016}).
\newblock


\bibitem[\protect\citeauthoryear{Dieleman, Schl{\"u}ter, Raffel, Olson,
  S{\o}nderby, Nouri, Maturana, Thoma, Battenberg, Kelly, Fauw, Heilman,
  diogo149, McFee, Weideman, takacsg84, peterderivaz, Jon, instagibbs, Rasul,
  CongLiu, Britefury, and Degrave}{Dieleman et~al\mbox{.}}{2015}]%
        {lasagne}
\bibfield{author}{\bibinfo{person}{Sander Dieleman}, \bibinfo{person}{Jan
  Schl{\"u}ter}, \bibinfo{person}{Colin Raffel}, \bibinfo{person}{Eben Olson},
  \bibinfo{person}{S{\o}ren~Kaae S{\o}nderby}, \bibinfo{person}{Daniel Nouri},
  \bibinfo{person}{Daniel Maturana}, \bibinfo{person}{Martin Thoma},
  \bibinfo{person}{Eric Battenberg}, \bibinfo{person}{Jack Kelly},
  \bibinfo{person}{Jeffrey~De Fauw}, \bibinfo{person}{Michael Heilman},
  \bibinfo{person}{diogo149}, \bibinfo{person}{Brian McFee},
  \bibinfo{person}{Hendrik Weideman}, \bibinfo{person}{takacsg84},
  \bibinfo{person}{peterderivaz}, \bibinfo{person}{Jon},
  \bibinfo{person}{instagibbs}, \bibinfo{person}{Dr.~Kashif Rasul},
  \bibinfo{person}{CongLiu}, \bibinfo{person}{Britefury}, {and}
  \bibinfo{person}{Jonas Degrave}.} \bibinfo{year}{2015}\natexlab{}.
\newblock \bibinfo{title}{Lasagne: First release.}
\newblock   (\bibinfo{date}{Aug.} \bibinfo{year}{2015}).
\newblock
\showDOI{%
\url{http://dx.doi.org/10.5281/zenodo.27878}}


\bibitem[\protect\citeauthoryear{Guadarrama and Silberman}{Guadarrama and
  Silberman}{2016}]%
        {slim}
\bibfield{author}{\bibinfo{person}{Sergio Guadarrama} {and}
  \bibinfo{person}{Nathan Silberman}.} \bibinfo{year}{2016}\natexlab{}.
\newblock \bibinfo{title}{TF Slim}.
\newblock
  \bibinfo{howpublished}{\url{https://github.com/tensorflow/tensorflow/tree/master/tensorflow/contrib/slim}}.
    (\bibinfo{year}{2016}).
\newblock


\bibitem[\protect\citeauthoryear{Jia, Shelhamer, Donahue, Karayev, Long,
  Girshick, Guadarrama, and Darrell}{Jia et~al\mbox{.}}{2014}]%
        {caffe}
\bibfield{author}{\bibinfo{person}{Yangqing Jia}, \bibinfo{person}{Evan
  Shelhamer}, \bibinfo{person}{Jeff Donahue}, \bibinfo{person}{Sergey Karayev},
  \bibinfo{person}{Jonathan Long}, \bibinfo{person}{Ross Girshick},
  \bibinfo{person}{Sergio Guadarrama}, {and} \bibinfo{person}{Trevor Darrell}.}
  \bibinfo{year}{2014}\natexlab{}.
\newblock \showarticletitle{Caffe: Convolutional Architecture for Fast Feature
  Embedding}. In \bibinfo{booktitle}{{\em Proceedings of the 22Nd ACM
  International Conference on Multimedia}} {\em (\bibinfo{series}{MM '14})}.
  \bibinfo{publisher}{ACM}, \bibinfo{address}{New York, NY, USA},
  \bibinfo{pages}{675--678}.
\newblock
\showISBNx{978-1-4503-3063-3}
\showDOI{%
\url{http://dx.doi.org/10.1145/2647868.2654889}}


\bibitem[\protect\citeauthoryear{Meng, Bradley, Yavuz, Sparks, Venkataraman,
  Liu, Freeman, Tsai, Amde, Owen, Xin, Xin, Franklin, Zadeh, Zaharia, and
  Talwalkar}{Meng et~al\mbox{.}}{2016}]%
        {mllib}
\bibfield{author}{\bibinfo{person}{Xiangrui Meng}, \bibinfo{person}{Joseph
  Bradley}, \bibinfo{person}{Burak Yavuz}, \bibinfo{person}{Evan Sparks},
  \bibinfo{person}{Shivaram Venkataraman}, \bibinfo{person}{Davies Liu},
  \bibinfo{person}{Jeremy Freeman}, \bibinfo{person}{DB Tsai},
  \bibinfo{person}{Manish Amde}, \bibinfo{person}{Sean Owen},
  \bibinfo{person}{Doris Xin}, \bibinfo{person}{Reynold Xin},
  \bibinfo{person}{Michael~J. Franklin}, \bibinfo{person}{Reza Zadeh},
  \bibinfo{person}{Matei Zaharia}, {and} \bibinfo{person}{Ameet Talwalkar}.}
  \bibinfo{year}{2016}\natexlab{}.
\newblock \showarticletitle{MLlib: Machine Learning in Apache Spark}.
\newblock \bibinfo{journal}{{\em J. Mach. Learn. Res.\/}} \bibinfo{volume}{17},
  \bibinfo{number}{1} (\bibinfo{date}{Jan.} \bibinfo{year}{2016}),
  \bibinfo{pages}{1235--1241}.
\newblock
\showISSN{1532-4435}
\showURL{%
\url{http://dl.acm.org/citation.cfm?id=2946645.2946679}}


\bibitem[\protect\citeauthoryear{Pedregosa, Varoquaux, Gramfort, Michel,
  Thirion, Grisel, Blondel, Prettenhofer, Weiss, Dubourg, Vanderplas, Passos,
  Cournapeau, Brucher, Perrot, and Duchesnay}{Pedregosa et~al\mbox{.}}{2011}]%
        {sklearn}
\bibfield{author}{\bibinfo{person}{Fabian Pedregosa}, \bibinfo{person}{Ga\"{e}l
  Varoquaux}, \bibinfo{person}{Alexandre Gramfort}, \bibinfo{person}{Vincent
  Michel}, \bibinfo{person}{Bertrand Thirion}, \bibinfo{person}{Olivier
  Grisel}, \bibinfo{person}{Mathieu Blondel}, \bibinfo{person}{Peter
  Prettenhofer}, \bibinfo{person}{Ron Weiss}, \bibinfo{person}{Vincent
  Dubourg}, \bibinfo{person}{Jake Vanderplas}, \bibinfo{person}{Alexandre
  Passos}, \bibinfo{person}{David Cournapeau}, \bibinfo{person}{Matthieu
  Brucher}, \bibinfo{person}{Matthieu Perrot}, {and}
  \bibinfo{person}{\'{E}douard Duchesnay}.} \bibinfo{year}{2011}\natexlab{}.
\newblock \showarticletitle{Scikit-learn: Machine Learning in Python}.
\newblock \bibinfo{journal}{{\em J. Mach. Learn. Res.\/}}  \bibinfo{volume}{12}
  (\bibinfo{date}{Nov.} \bibinfo{year}{2011}), \bibinfo{pages}{2825--2830}.
\newblock
\showISSN{1532-4435}
\showURL{%
\url{http://dl.acm.org/citation.cfm?id=1953048.2078195}}


\bibitem[\protect\citeauthoryear{Tokui, Oono, Hido, and Clayton}{Tokui
  et~al\mbox{.}}{2015}]%
        {chainer}
\bibfield{author}{\bibinfo{person}{Seiya Tokui}, \bibinfo{person}{Kenta Oono},
  \bibinfo{person}{Shohei Hido}, {and} \bibinfo{person}{Justin Clayton}.}
  \bibinfo{year}{2015}\natexlab{}.
\newblock \showarticletitle{Chainer: a Next-Generation Open Source Framework
  for Deep Learning}. In \bibinfo{booktitle}{{\em Proceedings of Workshop on
  Machine Learning Systems (LearningSys) in The Twenty-ninth Annual Conference
  on Neural Information Processing Systems (NIPS)}}.
\newblock
\showURL{%
\url{http://learningsys.org/papers/LearningSys_2015_paper_33.pdf}}


\bibitem[\protect\citeauthoryear{van Merri{\"{e}}nboer, Bahdanau, Dumoulin,
  Serdyuk, Warde{-}Farley, Chorowski, and Bengio}{van Merri{\"{e}}nboer
  et~al\mbox{.}}{2015}]%
        {blocksandfuel}
\bibfield{author}{\bibinfo{person}{Bart van Merri{\"{e}}nboer},
  \bibinfo{person}{Dzmitry Bahdanau}, \bibinfo{person}{Vincent Dumoulin},
  \bibinfo{person}{Dmitriy Serdyuk}, \bibinfo{person}{David Warde{-}Farley},
  \bibinfo{person}{Jan Chorowski}, {and} \bibinfo{person}{Yoshua Bengio}.}
  \bibinfo{year}{2015}\natexlab{}.
\newblock \showarticletitle{Blocks and Fuel: Frameworks for deep learning}.
\newblock \bibinfo{journal}{{\em CoRR\/}}  \bibinfo{volume}{abs/1506.00619}
  (\bibinfo{year}{2015}).
\newblock
\showURL{%
\url{http://arxiv.org/abs/1506.00619}}


\end{thebibliography}

\end{document}